\documentclass[prl,twocolumn,amsmath,superscriptaddress,amssymb]{revtex4-2}

\usepackage{graphicx}
\usepackage{dcolumn}
\usepackage{bm}
\usepackage{amssymb}
\usepackage{amsmath}
\usepackage{wasysym}
\usepackage{color}
\usepackage{times}

\usepackage{placeins}
\usepackage{epstopdf}
\usepackage{upgreek}
\usepackage{hyperref}

\usepackage[normalem]{ulem}

\makeatletter
\renewcommand*\env@matrix[1][c]{\hskip -\arraycolsep
  \let\@ifnextchar\new@ifnextchar
  \array{*\c@MaxMatrixCols #1}}
\makeatother
\begin{document}

\title{Topological band inversion in HgTe(001): surface and bulk signatures from photoemission}
\author{Raphael C. Vidal}\affiliation{Experimentelle Physik VII and W\"urzburg-Dresden Cluster of Excellence ct.qmat, Universit\"at W\"urzburg, Am Hubland, D-97074 W\"urzburg, Germany, EU}
\author{Giovanni Marini}\affiliation{Department of Physical and Chemical Sciences \& SPIN-CNR, University of L'Aquila, Italy, EU}
\author{Lukas Lunczer}\affiliation{Experimental physics III, Universit\"at W\"urzburg, Am Hubland, D-97074 W\"urzburg, Germany, EU}
\author{Simon Moser}\affiliation{Advanced Light Source, Lawrence Berkeley National Laboratory, Berkeley, CA 94720, USA}\affiliation{Experimental physics IV and W\"urzburg-Dresden Cluster of Excellence ct.qmat, Universit\"at W\"urzburg, Am Hubland, D-97074 W\"urzburg, Germany, EU}
\author{Lena F\"urst}\affiliation{Experimental physics III, Universit\"at W\"urzburg, Am Hubland, D-97074 W\"urzburg, Germany, EU}
\author{Chris Jozwiak}\affiliation{Advanced Light Source, Lawrence Berkeley National Laboratory, Berkeley, CA 94720, USA}
\author{Aaron Bostwick}\affiliation{Advanced Light Source, Lawrence Berkeley National Laboratory, Berkeley, CA 94720, USA}
\author{Eli Rotenberg}\affiliation{Advanced Light Source, Lawrence Berkeley National Laboratory, Berkeley, CA 94720, USA}
\author{Charles Gould}\affiliation{Experimental physics III, Universit\"at W\"urzburg, Am Hubland, D-97074 W\"urzburg, Germany, EU}
\author{Hartmut Buhmann}\affiliation{Experimental physics III, Universit\"at W\"urzburg, Am Hubland, D-97074 W\"urzburg, Germany, EU}
\author{Wouter Beugeling}\affiliation{Experimental physics III, Universit\"at W\"urzburg, Am Hubland, D-97074 W\"urzburg, Germany, EU}
\author{Giorgio Sangiovanni}\affiliation{Institut f\"ur Theoretische Physik und Astrophysik and W\"urzburg-Dresden Cluster of Excellence ct.qmat, Universit\"at W\"urzburg, 97074 W\"urzburg, Germany, EU}
\author{Domenico Di Sante}\affiliation{Theoretische Physik I, Universit\"at W\"urzburg, Am Hubland, D-97074 W\"urzburg, Germany, EU}\affiliation{Department of Physics and Astronomy, University of Bologna, 40127 Bologna, Italy, EU}\affiliation{Center for Computational Quantum Physics, Flatiron Institute, 162 5th Avenue, New York, New York 10010, USA}
\author{Gianni Profeta}\affiliation{Department of Physical and Chemical Sciences \& SPIN-CNR, University of L'Aquila, Italy, EU}
\author{Laurens W. Molenkamp}\affiliation{Experimental physics III, Universit\"at W\"urzburg, Am Hubland, D-97074 W\"urzburg, Germany, EU}
\author{Hendrik Bentmann}\email{Hendrik.Bentmann@physik.uni-wuerzburg.de}\thanks{Present address: Center for Quantum Spintronics, Department of Physics, Norwegian University of Science and Technology, NO-7491 Trondheim, Norway}\affiliation{Experimentelle Physik VII and W\"urzburg-Dresden Cluster of Excellence ct.qmat, Universit\"at W\"urzburg, Am Hubland, D-97074 W\"urzburg, Germany, EU}
\author{Friedrich Reinert}\affiliation{Experimentelle Physik VII and W\"urzburg-Dresden Cluster of Excellence ct.qmat, Universit\"at W\"urzburg, Am Hubland, D-97074 W\"urzburg, Germany, EU}

\date{\today}
\begin{abstract}
HgTe is a versatile topological material and has enabled the realization of a variety of topological states, including two- and three-dimensional (3D) topological insulators and topological semimetals. Nevertheless, a quantitative understanding of its electronic structure remains challenging, in particular due to coupling of the Te 5$p$-derived valence electrons to Hg 5$d$ core states at shallow binding energy. We present a joint experimental and theoretical study of the electronic structure in strained HgTe(001) films in the 3D topological-insulator regime, based on angle-resolved photoelectron spectroscopy and density functional theory. The results establish detailed agreement in terms of (i) electronic band dispersions and orbital symmetries, (ii) surface and bulk contributions to the electronic structure, and (iii) the importance of Hg 5$d$ states in the valence-band formation. Supported by theory, our experiments directly image the paradigmatic band inversion in HgTe, underlying its non-trivial band topology. 
\end{abstract}
\maketitle

Topological band theory marks a milestone in condensed matter physics and has established a deeper understanding of electronic structure in crystalline solids \cite{Kane:05,Fu:07,Moore:07,Qi:11}. It has led to the discovery of a variety of topologically non-trivial states of matter, including two- and three-dimensional (2D and 3D) topological insulators (TI) \cite{konig2007quantum,Hasan:10}, topological crystalline insulators \cite{tjernberg:12}, magnetic TI \cite{chang:13,otrokov:19}, as well as Dirac and Weyl semimetals \cite{Armitage:18}. These topological states form the basis for unusual electron transport phenomena, such as the quantum spin Hall effect in 2D TI \cite{konig2007quantum}, the quantum anomalous Hall effect in magnetic TI \cite{chang:13}, and the chiral anomaly in Weyl semimetals \cite{Armitage:18}. Although, according to recent theoretical predictions, one may, in principle, expect an abundance of topological materials in nature \cite{vergniory:19}, the number of systems that allow for the experimental observation and control of these phenomena is still very limited. The compound HgTe constitutes one such paradigmatic topological material. Through growth of epitaxial films, a variety of topological regimes were realized in HgTe in dependence of film thickness and lattice strain. For instance, films of HgTe allowed for the observation of the quantum spin Hall effect \cite{konig2007quantum} as well as, more recently, a chiral-anomaly driven negative magnetoresistance \cite{mahler2019interplay}, and signatures of topological superconductivity \cite{bocquillon_gapless_2017} and of Majorana quasi-particles \cite{ren_topological_2019}.   

Despite the aforementioned importance of HgTe in the field of topological materials, a quantitative understanding of its electronic band structure still remains difficult \cite{Nicklas:11,Sakuma:11,Kotani:11}. The origin of the non-trivial topology in HgTe is a band inversion around the $\Gamma$-point of the Brillouin zone (BZ), where states of Hg 6$s$ orbital character and states of Te 5$p$ character acquire an inverted energetic order around the band gap, giving rise to an inverted gap \cite{bernevig_quantum_2006}. By now, similar inversions of energy levels have been recognized as a key signature of topologically non-trivial band structures \cite{Bansil:16}. While the general band-inversion mechanism in HgTe has been known for a long time \cite{Groves:63,Groves:67}, an accurate description of the band structure from first principles is challenging. A main reason for the difficulties lies in the presence of Hg 5$d$ semicore states at shallow binding energies that influence the $sp$ band structure and the inverted band gap via $p-d$ interaction \cite{Zunger:88,Fleszar:05}. Despite previous photoemission works \cite{Manzke:00,Hasan:15}, systematic comparison of experiment and theory has remained largely limited to band positions at the $\Gamma$-point, as obtained from optical measurements \cite{Groves:67,Chadi:72,moritani_electroreflectance_1973}, which, however, provide no momentum-dependent and no surface-sensitive information.

\begin{figure*}[t]
\begin{center}
\includegraphics[width=.95\linewidth]{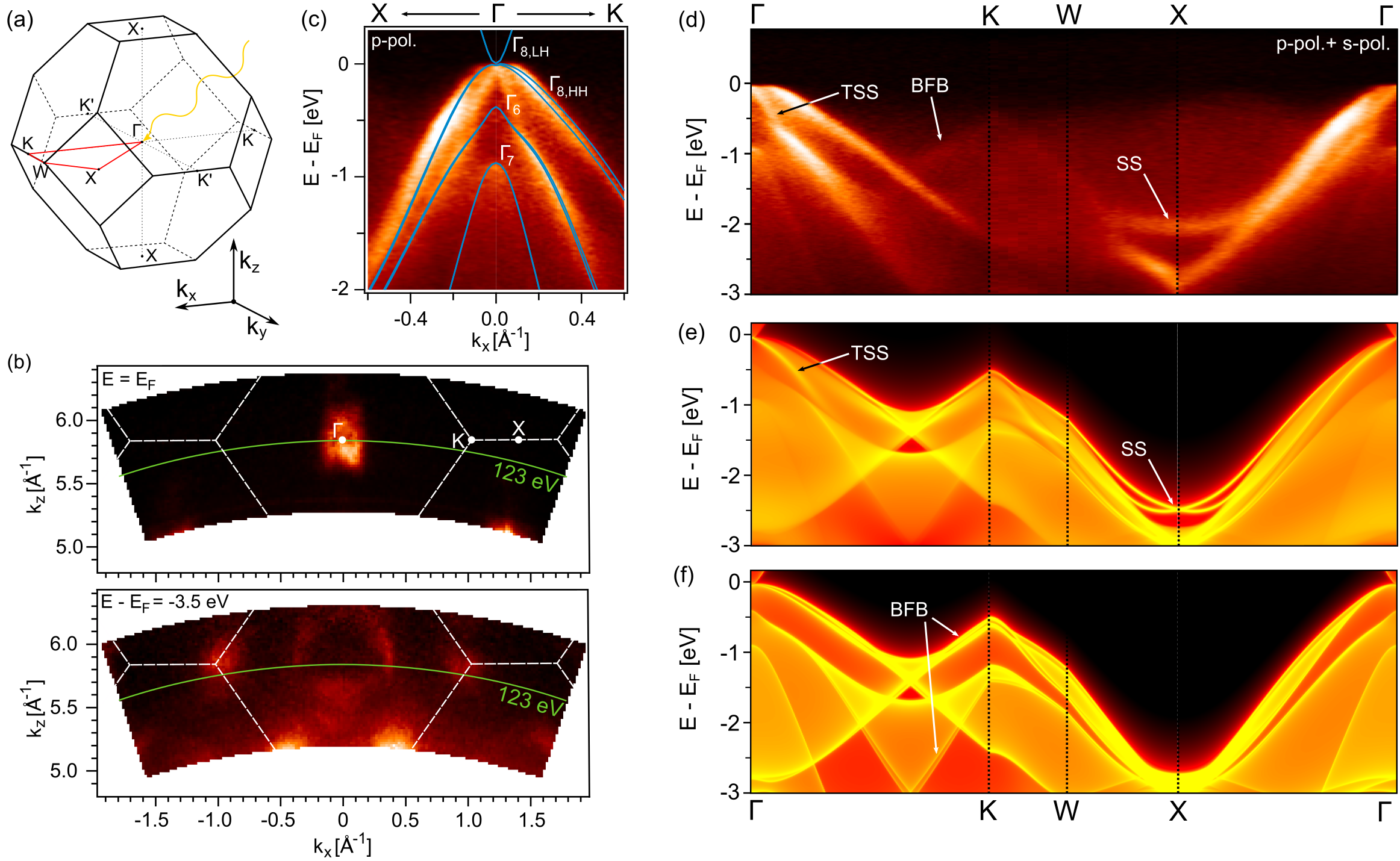}
\caption{(color online) (a) Bulk Brillouin zone with high-symmetry points. The $k$-space path for the panels (d)-(f) is indicated in red. (b) $k_{x}$-$k_{z}$ ARPES data sets obtained from $h\nu$-dependent measurements. (c) ARPES data set for HgTe(001) around the $\Gamma$-point ($h\nu = 123\,$eV) and corresponding calculation of the bulk band structure. (d) ARPES data along high-symmetry directions. For maximal visibility we plot the sum of data sets for $p$- and $s$-polarized light (cf.~Fig.~3). (e)-(f) Calculations of (001)-projected surface and bulk spectral functions.}
\label{fig1}
\end{center}
\end{figure*}

In the present work, we report on the electronic structure of HgTe(001) films in the 3D TI regime. Employing photon-energy- and polarization-dependent angle-resolved photoelectron spectroscopy (ARPES) we determine dispersions and orbital symmetries of the $sp$-derived valence bands and disentangle bulk and surface states. Our experiments also unveil an itinerant character of the Hg 5$d$ states, as evidenced by a finite band dispersion and sub-band splittings of the $d_{\frac{5}{2}}$ and $d_{\frac{3}{2}}$ manifolds. The experimental results are in detailed agreement with calculations based on density functional theory (DFT) performed using the HSE06 hybrid functional on top of local-density approximation for exchange and correlation. For other approximations the agreement is considerably inferior. Our findings establish a benchmark for the investigation of topological phenomena in HgTe-based systems and in related material classes, where analogous topological band inversions have been predicted \cite{chadov_tunable_2010,lin_half-heusler_2010}.   



The HgTe films were grown by molecular beam epitaxy on CdTe substrates (see Fig.~S1 for the detailed layer stacking). The samples were either grown on  with a protective amorphous Te/Ti layer or transferred directly with an UHV suitcase to the ARPES measurement setup. We conducted ARPES experiments at the MAESTRO endstation ($\upmu$ARPES setup) at beamline 7 of the Advanced Light Source (ALS) and at a high-resolution ARPES setup in the laboratory in Würzburg. Band-structure calculations were performed based on DFT. Experimental and theoretical details are given in the  Supplemental material.

\begin{figure}[t]
\begin{center}
\includegraphics[width=.95\linewidth]{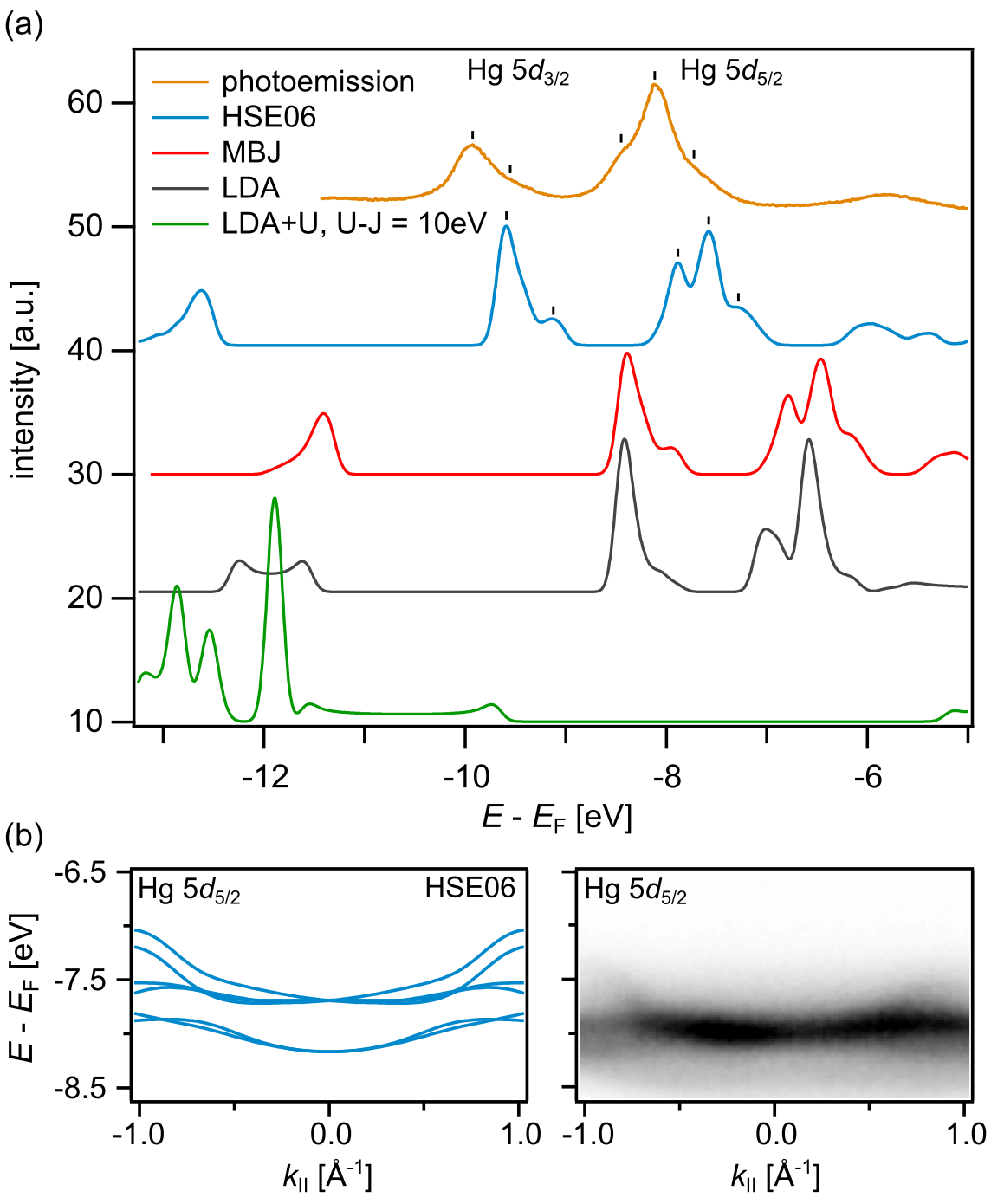}
\caption{(color online) (a) Angle-integrated photoemission data of the Hg 5$d$ states in comparison to DFT calculations based on different exchange-correlation functionals. (b) Calculated and measured band dispersion of the Hg 5$d_{\frac{5}{2}}$ states.}
\label{fig2}
\end{center}
\end{figure}

\begin{figure*}[t]
\begin{center}
\includegraphics[width=.95\linewidth]{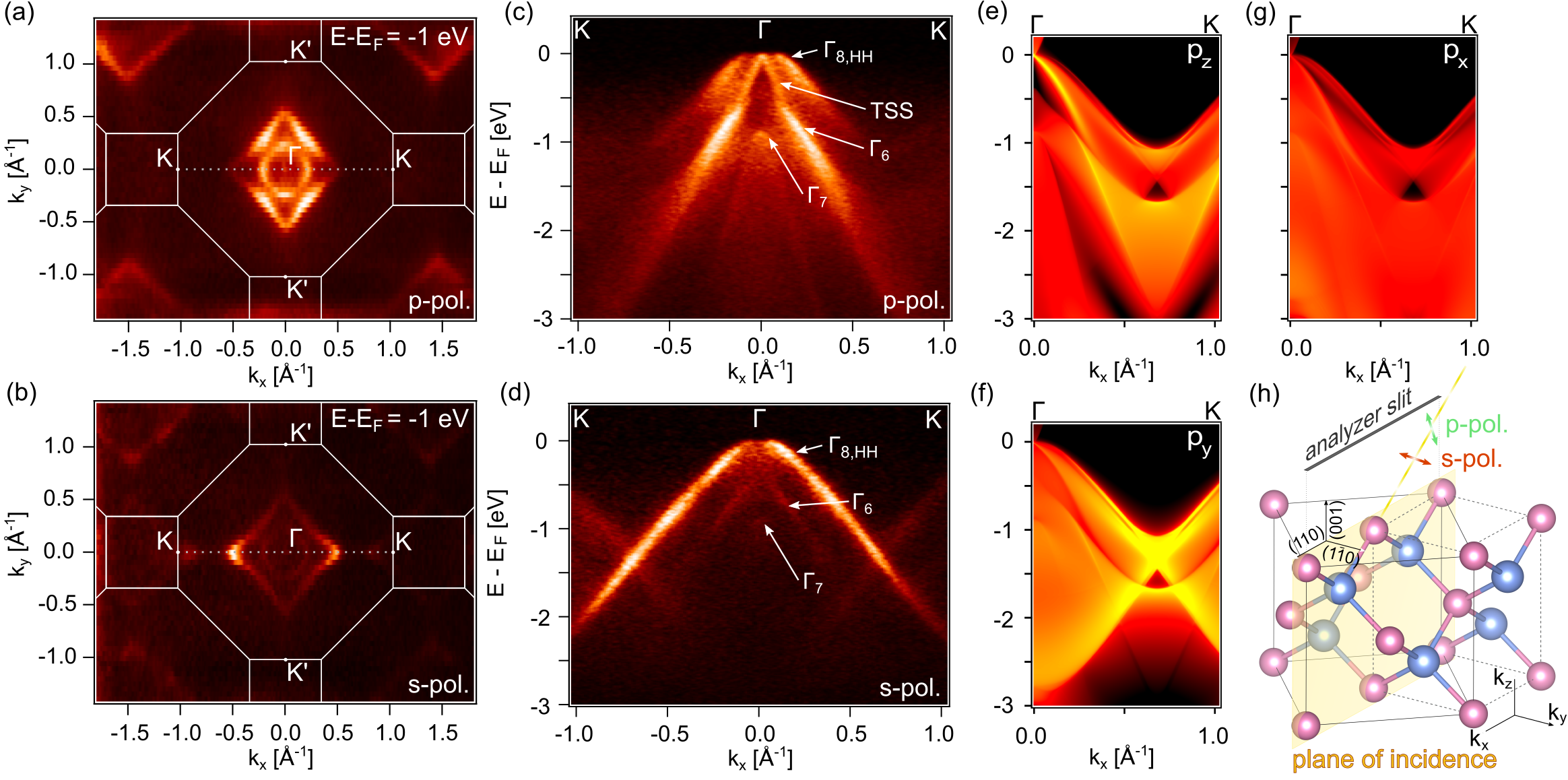}
\caption{(color online) (a)-(b) Constant-energy ARPES data sets obtained with $p$- and $s$-polarized light. The boundaries of the bulk BZ are indicated. (c)-(d) ARPES data along the K-$\Gamma$-K-direction for $p$- and $s$-polarized light. (e)-(g) Calculations of the (001)-projected surface spectral function along $\Gamma K$ projected on different Te 5$p$ orbitals. (h) Schematic of the experimental geometry. The data sets in all panels were taken with $h\nu = 123\,$eV.}
\label{fig3}
\end{center}
\end{figure*}


We collected $h\nu$-dependent ARPES data to identify the high-symmetry planes in the the 3D band structure of HgTe [Fig.~\ref{fig1}(a)-(b) and Fig. S6]. The $\Gamma$-point of the bulk BZ is reached in normal emission at a photon energy of approximately $h\nu =\,$123$\,$eV, consistent with an inner potential of $V_0=\,$10.3~eV. Fig.~1(c) shows a corresponding ARPES data set together with the calculated bulk band structure along high-symmetry directions. A comparison over a larger energy range is shown in Fig.~S5 of the Supplementary material. We identify the $\Gamma_8$, $\Gamma_6$ and $\Gamma_7$ valence bands. An additional feature, which is not found in the bulk calculation, is attributed to the topological surface state (TSS). As discussed below in more detail (cf. Fig.~4), the TSS disperses linearly from the Fermi level to higher binding energies where it merges with the $\Gamma_6$ bulk band.  

To disentangle surface and bulk contributions to the electronic structure, we compare ARPES data to calculations of the (001)-projected spectral functions along a high-symmetry path in the BZ  [Fig.~\ref{fig1}(d)-(f)]. Considering bulk- and surface-projected spectral weights we identify two main surface features: the TSS near the $\Gamma$-point and another surface state (SS) at the $X$-point, which coincides with the $\bar{K}$-point of the surface BZ. To our knowledge, the surface state at $X$ has not been identified before. Both features are confirmed by the experimental data. All other features are attributed primarily to the bulk, consistent with the decent match of experiment and bulk calculation in Fig.~\ref{fig1}(c). Another effect of the (001)-projection is the appearance of backfolded bands (BFB) in the first bulk BZ. Along the $\Gamma K$ direction these BFB cross the primary bands at the boundary of the surface BZ, i.e. at the $\bar{J}$-point. The BFB are also discerned in the experimental data, albeit with comparably low spectral weight. 

The level of agreement between measured and calculated electronic structure, established in Fig.~1, becomes significantly compromised for other approximations of exchange and correlation, as seen from a detailed comparison of the experimental data to calculations based on LDA, LDA+$U$, and the MBJ hybrid functional in the Supplementary Figs.~S3 and S4. The importance of exchange-correlation effects in HgTe has been attributed to $p-d$ interaction of the $sp$ valence bands with the shallow Hg 5$d$ states \cite{Zunger:88,Fleszar:05}, which we address in the following. 

Fig.~\ref{fig2}(a) compares angle-integrated photoemission data of the Hg 5$d$ states to calculations based on different exchange-correlation functionals. The best match to the experimental binding energy is obtained with the HSE06 functional which, accounting for a percentage of the exact Fock exchange, partially corrects the self-interaction inherently associated to local DFT functionals. Indeed, LDA and MBJ severely underestimate the 5$d$ binding energy, while LDA+$U$ gives rise to strong overestimations for reasonable values of $U$. Our experimental data further reveals a splitting of the $d_{\frac{5}{2}}$ and $d_{\frac{3}{2}}$ peaks into three and two components, respectively. This splitting is again nicely captured by our HSE06-based calculation and thus can be attributed to an initial-state effect in the 5$d$ levels. As seen in the momentum-dependent calculation in Fig.~\ref{fig2}(b), this fine structure arises from a subband-splitting and a finite band dispersion, both evidencing a participation of the 5$d$ states in the valence-band formation. Angle-resolved measurements confirm the dispersive character of the 5$d$ states with a total band width of approximately 1~eV for the $d_{\frac{5}{2}}$ level [Fig.~\ref{fig2}c and Figs.~S5-S7]. Supported by theory, our experiments thus establish evidence for significant $p-d$ hybridization in HgTe.

Our calculations further show a spin splitting of the $\Gamma_8$-band along $\Gamma X$ ([110] direction), which remains unresolved in our measurements. It has been proposed that the linear term of the splitting, close to the $\Gamma$-point, arises from an admixture of $d$ orbital character and can be written as $\Delta k_x =  \frac{3}{2}\sqrt{3}C k$, where the coefficient $C$ is given by  \cite{dresselhaus1955spin,cardona1986terms,cardona1988relativistic}: 
\begin{equation}
C = -A \frac{\Delta_{d,c}}{E(\Gamma_8)-E_{d,c}}.
\label{eq1}
\end{equation}
$E(\Gamma_8)$ is the top of the $\Gamma_8$-band, and $\Delta_{d,c}$ is the spin-orbit splitting and $E_{d,c}$ the energy of the Hg 5$d$ levels. The constant $A$ has been estimated to $350\,\mathrm{meV}\mathrm{{\AA}}$ for II-VI compounds \citep{cardona1986terms}. From our direct band calculation we estimate $C \sim -70\,\mathrm{meV}\mathrm{{\AA}}$ (Fig.~S7), while using Eq.~1 we obtain $C \sim -80\,\mathrm{meV}\mathrm{{\AA}}$ for experimental values and $C\sim -87.5\,\mathrm{meV}\mathrm{{\AA}}$ for theoretical values. The reasonable match supports the validity of Eq.~(1) and, therefore, indicates an influence of $p-d$ interaction on the spin splitting of the top-most valence band states. 

The spin splitting arises from bulk inversion asymmetry (BIA), which quantifies the difference between the monoatomic diamond structure (similar to $\alpha$-Sn) and the HgTe zincblende structure and can play a crucial role for transport phenomena in HgTe. For example, in compressively strained HgTe, Weyl points arise as a result of BIA, in contrast to Dirac fermions present without inversion asymmetry \cite{Ruan2016,mahler2019interplay}. In HgTe quantum wells, an even-odd effect in the length of the Hall conductance plateaus could be explained by BIA \cite{Shamim2020}.
For quantitative analysis of these phenomena, one commonly uses $\mathbf{k}\cdot\mathbf{p}$-theory \cite{Winkler2003}, but the strength of the BIA couplings of HgTe is poorly known.
In order to provide an estimate from our present results, we fit a $4\times 4$ Luttinger model \cite{Ruan2016} with a linear BIA term  $\mathcal{H}_{\mathrm{BIA}} = \alpha[k_x\{J_x,J_y^2-J_z^2\} + \text{c.p.}]$  (analogous to $\frac{3}{2}\sqrt{3}C k$ above) to the band structure obtained from the HSE06 functional. We obtain $\alpha = 70\pm 20$ meV\,\AA. More details are provided in Supplementary Section IV.

We proceed by examining the orbital composition of the valence bands, which plays a crucial role for the topological properties. 
We exploit selection rules imposed by the dipole approximation of the photoemission matrix element. A sketch of the experimental geometry is shown in Fig. \ref{fig3}(g). The plane of light incidence ($xz$ plane) is aligned with the mirror-symmetric \{110\}-plane, implying that, for emission within the plane of light incidence, linearly $s$-polarized light couples exclusively to odd orbitals and $p$-polarized light to even orbitals. A strong influence of the light polarization on the intensity distributions is directly apparent from the constant-energy maps in Figs.~\ref{fig3}(a)-(b). 

Focusing on the intensity within the plane of light incidence in Figs.~\ref{fig3}(c)-(d), we find that $s$-polarized light predominantly excites the $\Gamma_8$-band. This is in line with our calculations in Figs.~\ref{fig3}(e)-(g) assigning a main contribution from odd $p_y$ orbitals to this band. All other features are largely suppressed for $s$-polarization with the exception of minor contributions from the $\Gamma_6$-band at small but finite $k_x$, consistent with the calculations showing a sligthly enhanced $p_y$ character at similar $k_x$. Comparing the momentum dependences of the ARPES spectral weight for $p$-polarized light and of the projections on even $p_x$ and $p_z$ orbitals, we conclude that the measurement mainly reflects $p_z$-derived states, including the TSS as well as parts of the $\Gamma_8$- and $\Gamma_6$-bands.  

\begin{figure}[t]
\begin{center}
\includegraphics[width=.95\linewidth]{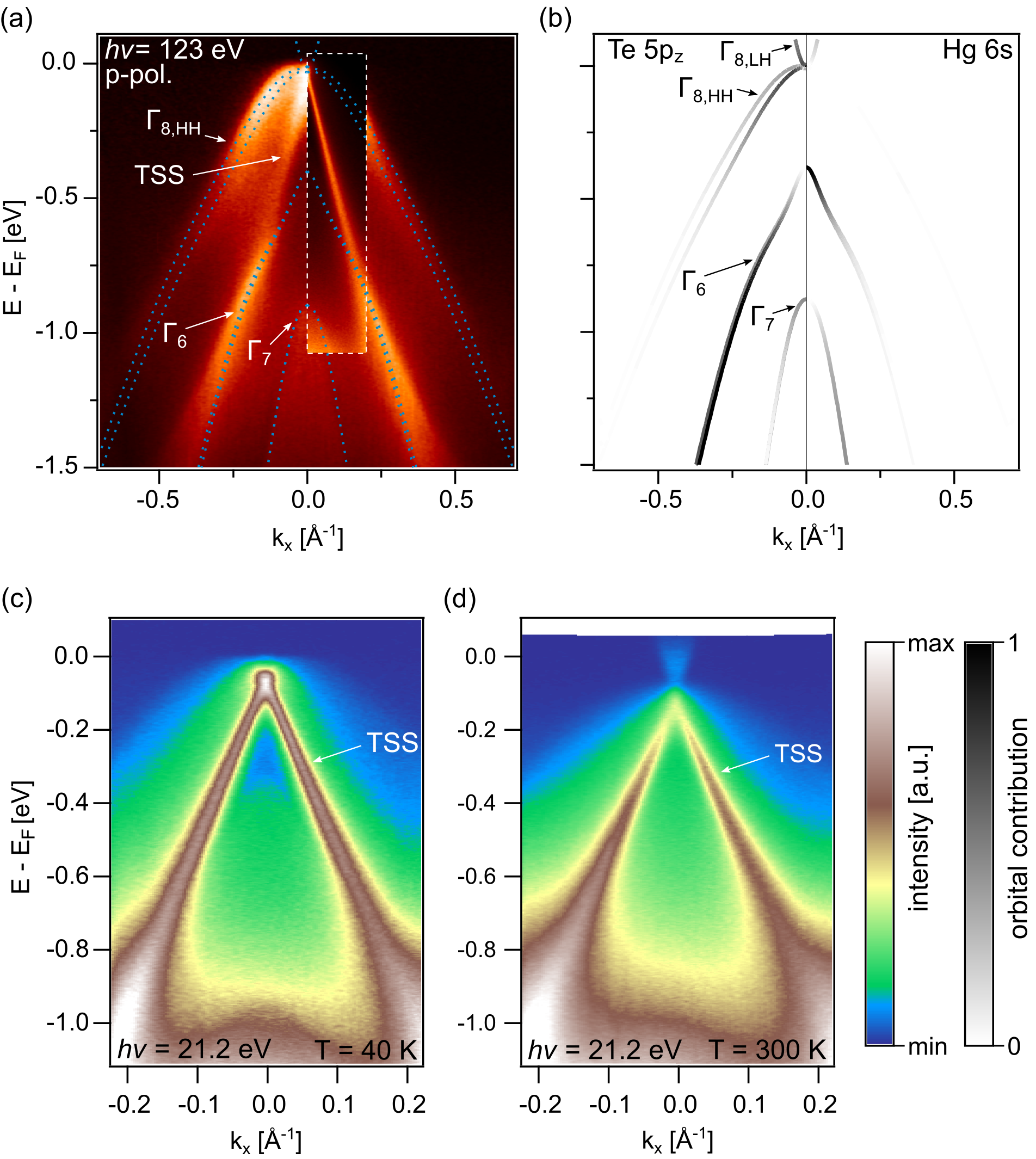}
\caption{(color online) (a) ARPES data near the $\Gamma$-point taken with $h\nu = 123\,$eV and p-polarized light. The inset shows ARPES data obtained with with He $\mathrm{I}_\alpha$-line ($h\nu = 21.2\,$eV), for which the cross section of the topological surface state is much larger than for the bulk bands. (b) Calculated bulk band structure projected on Te 5$p_z$ and Hg 6$s$ orbitals. (c)-(d) ARPES data sets measured at (b) $\mathrm{T} = 40\,$K and (c) $300\,$K ($h\nu = 21.2\,$eV).}
\label{fig4}
\end{center}
\end{figure}

A more detailed analysis near the $\Gamma$-point is shown in Fig.~\ref{fig4}. The measured intensity distribution for p-polarized light in Fig.~\ref{fig4}(a) displays a similar momentum dependence as the calculated $p_z$-projected weight of the bulk bands [Fig.~\ref{fig4}(b)]. In particular, drops in $p_z$ orbital character towards large wave vectors for the $\Gamma_8$-band and towards small wave vectors for the $\Gamma_6$-band are reflected in the experimental data, where corresponding reductions of spectral weight are observed. As seen in the calculations in Fig.~\ref{fig4}(b), the momentum-dependent change in orbital contribution for the $\Gamma_6$-band is directly related to the band inversion, which gives rise to a dominating Hg 6$s$ character near $\Gamma$ at the expense of Te $p_z$ character. Due to the low cross section for $s$ orbitals the $\Gamma_6$-band is suppressed near $\Gamma$. The inset in Fig.~\ref{fig4}(a) shows data of the TSS taken at $h\nu = 21.2\,$eV, for which it is the dominating spectral feature [cf. Figs.~\ref{fig4}(c)-(d)]. It is evident that the TSS merges into the $\Gamma_6$-band near wave vectors where the latter strongly drops in intensity, i.e. where the orbital character changes from Te $p_z$ to Hg 6$s$. These observations provide direct evidence for the correlation between bulk band inversion and non-trivial surface electronic structure.

To obtain information about the conduction band we collected ARPES data at elevated temperature [Figs.~\ref{fig4}(c)-(d)]. A pronounced shift in energy is observed when comparing data at $300\,$K and at $40\,$K. At room temperature the bottom of the conduction band lies approximately 50~meV below the Fermi level, while it is not observed up to $E_F$ at low temperature.
According to our calculations, the energy shift arises from the strong density of states asymmetry between the top of the valence and
the bottom of the conduction band (Fig.S10). As a result of the increased temperature, the chemical potential is
shifted by 55 meV, which is in agreement with the experiment.                  

In summary, we present a comprehensive experimental and theoretical investigation of the electronic structure of HgTe(001) in the 3D TI regime. The experimental results are captured in detail by calculations based on density functional theory in a hybrid-functional approximation, indicating a crucial role of non-local exchange interactions in HgTe. Our findings provide direct spectroscopic evidence for the band inversion in the bulk electronic structure and for its relation to the existence of a topological surface state. Our results will enable an improved theoretical description of more complex topological phases in HgTe-based systems \cite{mahler2019interplay,bocquillon_gapless_2017,ren_topological_2019}, and may also facilitate the identification of topological states in related material classes. For instance, a variety of half-Heusler compounds were predicted to feature topological band inversions analogous to HgTe \cite{chadov_tunable_2010,lin_half-heusler_2010}, but experimental evidence is still scarce \cite{logan_observation_2016}. In these compounds $p-d$ interactions likely play a similarly important role as demonstrated here for HgTe.

\section{Acknowledgments}
We thank Eeshan Ketkar for critical reading of the manuscript. We acknowledge financial support from the DFG through SFB1170 'Tocotronics' (projects A01, A04, C05) and the W\"urzburg-Dresden Cluster of Excellence on Complexity and Topology in Quantum Matter -- \textit{ct.qmat} (EXC 2147, project-id 390858490). This research used resources of the Advanced Light Source (ALS), which is a DOE Office of Science User Facility under contract no. DE-AC02-05CH11231. The authors gratefully acknowledge the CINECA Supercomputing Center providing computational time through the ISCRA project. G. P. wishes to acknowledge financial support from the Italian Ministry for Research and Education through PRIN-2017 project ``Tuning and understanding quantum phases in 2D materials— Quantum 2D'' (IT-MIUR Grant No. 2017Z8TS5B). S.M. acknowledges support by the Swiss National Science Foundation under grant no. P300P2-171221.

\end{document}